\documentclass[aps,prb,twocolumn,showpacs]{revtex4}

\usepackage{amsmath}
\usepackage{graphicx}
\usepackage{multirow}
\usepackage{array}

\begin{document}
    \author{X. Fabr\`eges$^1$, I. Mirebeau$^1$, P. Bonville$^2$, S. Petit$^1$,
     G. Lebras-Jasmin$^2$, A. Forget$^2$, G. Andr\'e$^1$, S. Pailh\`es$^1$ }
    \date{01/04/2008}
    \affiliation{$^1$ Laboratoire L\'eon Brillouin, CEA-CNRS, CE-Saclay, 
     91191 Gif-sur-Yvette,France}
    \affiliation{$^2$ DSM/Service de Physique de l'Etat Condens\'e, CEA-CNRS, 
     CE-Saclay, 91191 Gif-Sur-Yvette, France}
    \title{Magnetic order in YbMnO$_3$ studied by Neutron Diffraction and 
     M\"ossbauer Spectroscopy}

    \begin{abstract}
        The magnetic ordering of the hexagonal multiferroic compound YbMnO$_3$
        has been studied between 100~K and 1.5~K by combining neutron powder
        diffraction, $^{170}$Yb M\"ossbauer spectroscopy and magnetization
        measurements. The Yb moments of the two crystallographic sites
        order at two different temperatures, the $4b$ site together with the 
	Mn moments (at $T_N \simeq$85~K) and the $2a$ site well below (at 
	3.5~K). The temperature dependences of the Mn and Yb moments are 
	explained within a molecular field model, showing that the $4b$ and 
	$2a$ sites order via Yb-Mn  and Yb-Yb interactions respectively. A 
	simple picture taking into account the local Mn environment of the 
	Rare earth R ($4b$) ion is proposed to couple R and Mn orders in 
	hexagonal RMnO$_3$ manganites. The nature and symmetry of the R-Mn 
	interactions yielding the R order are discussed.
    \end{abstract}
       \maketitle

    \section{Introduction}
        RMnO$_3$ manganites, where the R$^{3+}$ ions are either Y or a rare
        earth, belong to the multiferroic family, showing a coupling between
        magnetic and electric order parameters. This magneto-electric (M-E) 
	coupling opens the possibility of tuning the magnetization by an electric 
	field or vice-versa, with potential application for building new kinds of 
	electronic devices\cite{Fiebig_revue,Cheong_Nature}. Numerous experiments 
	in the RMnO$_3$ series probe the key role of their non collinear magnetic 
	structures, induced by frustrated magnetic interactions, in driving such 
	M-E coupling\cite{Park_Y}. The M-E coupling is also connected with a 
	spin-lattice coupling, recently observed both in structural\cite{Lee} and 
	dynamical\cite{Pimenov,Katsura,Sylvain} properties. Understanding the 
	microscopic characteristics of the magnetic order and the origin of the 
	magnetic interactions is a key issue for tuning their properties.

        The RMnO$_3$ compounds are divided in two classes according to the
        ionic radius of the rare earth. Orthorhombic structures are
        stabilized for  large ionic radius of the R$^{3+}$ ion (R=La-Dy),
        whereas hexagonal structures are stable at  small ionic radius (R=Y,
        Ho-Lu). In orthorhombic compounds where magnetic frustration arises from
        competing exchange interactions,
        ferroelectric and magnetic orders appear at the same temperature
        ($\sim$60~K), which suggests a strong M-E coupling\cite{Kimura}. In 
	hexagonal compounds, magnetic frustration arises from the lattice 
	geometry, since the triangular Mn lattice is frustrated for 
	antiferromagnetic (AF) first neighbor interactions. Here,
        the ferroelectric order occurs around 900~K, well above the magnetic
        ordering temperature ($\sim$80~K). The M-E coupling is evidenced for 
	instance by anomalies in the dielectric constant\cite{Huang_1997} or 
	specific heat \cite{Katsu_2001} at the magnetic transitions.

        In spite of numerous experiments, the complex magnetic structures
        of the RMnO$_3$ are still not fully understood. This is especially
        true for the hexagonal compounds, where  homometric
        configurations of the Mn moments in the triangular Mn lattice are
        extremely difficult to distinguish, and where the R magnetic order is
        intricate, owing to the presence of two different crystallographic  rare
        earth sites\cite{Munoz}.

        Here we present a detailed study of the magnetic order in YbMnO$_3$,
        using neutron diffraction, M\"ossbauer spectroscopy  on the isotope
        $^{170}$Yb and magnetization measurements. The magnetic structure of this 
	compound has not been studied up to now, although precise predictions 
	could be made from optical spectroscopy data\cite{Raman}.
        By combining  microscopic and macroscopic probes in a large temperature
        range (1.5~K$<$T$<$100~K), we provide a complete determination of the 
	magnetic structure.

        The T-dependence of the Mn and Yb moments derived from our experiments
        can be explained quantitatively within  the molecular  field model.
        This allows us to clarify the respective roles of Mn-Mn, R-Mn and R-R
        interactions in this compound. Using these results, we discuss the 
	symmetry and the possible origins of the magnetic R-Mn interactions, and 
	we propose a simple picture, which may hold for the whole hexagonal 
	family.

    \section{Experimental Details}
        A polycrystalline sample of YbMnO$_3$ was prepared as described in 
	Ref.\onlinecite{Alonso_2000} and characterized using X-Ray diffraction at 
	room temperature, showing that the sample is single phased. Magnetic 
	measurements were performed with a superconducting quantum interference 
	device (SQUID) magnetometer between 1.5~K and 300~K.

        M\"ossbauer spectroscopy on $^{170}$Yb realises $\gamma$-ray resonant
        absorption between the ground nuclear state, with spin $I_g$=0,
        and the first excited state, with spin $I_e$=2, at an energy of 84.3\,keV.
        We used a neutron activated Tm$^*$B$_{12}$ $\gamma$-ray source, mounted 
	on a triangular velocity electromagnetic drive. $^{170}$Yb M\"ossbauer 
	absorption spectra were recorded in the temperature range 1.35 - 50~K.

        The crystal structure was determined  by measuring a neutron powder
        diffraction (NPD) pattern at 300~K on the high resolution powder
        diffractometer 3T2 of the Laboratoire L\'eon Brillouin (LLB) at  Orph\'ee
        reactor, with an incident neutron wavelength $\lambda=1.2253$~\AA. The
        magnetic structure was studied by collecting NPD patterns at several
        temperatures, between 100~K (above the magnetic transition) and 1.5~K, on 
	the neutron diffractometer G4-1 of the LLB, with an incident neutron 
	wavelength of $2.4226$~\AA. Both crystal and magnetic structures were 
	refined using the Fullprof suite\cite{Rod_Carv_1993}.

    \section{Crystal Structure}
\begin{figure}
               \centering
               \includegraphics[width=8cm,height=5.5cm]{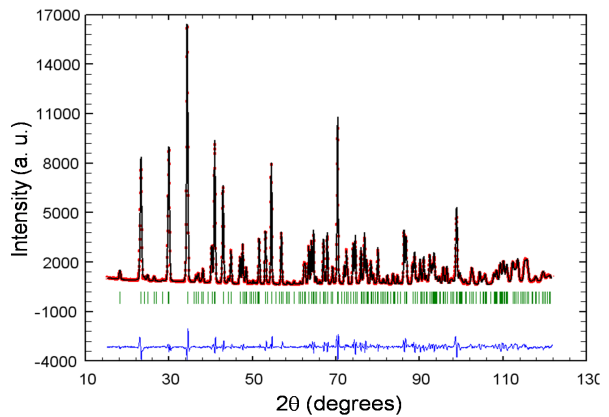}
               \caption{(Color online) Observed and Fullprof calculated NPD 
	pattern at room temperature. The Bragg reflections (tics), and the 
	difference between the observed and calculated patterns are plotted at 
	the bottom.}
               \label{3T2}
           \end{figure}

        The refined NPD pattern at 300 K is shown in Fig.\ref{3T2}. All Bragg 
	reflexions of the pattern can be indexed within the hexagonal space group 
	$P6_3cm$. The lattice constants at 300 K are $a=6.0629(1)$~\AA\ and 
	$c=11.3529(1)$~\AA.

        The refined atomic positions are reported in Table \ref{position}. They 
	are in good agreement with previous determinations from X-ray 
	diffraction\cite{Isobe_1991,Van_2001}. As already stated in 
	Ref.\onlinecite{Munoz_2000}, the lattice unit cell consists of 6 formula 
	units. Each Mn atom is surrounded by oxygen ions forming a MnO$_5$ 
	bipyramidal structure, with 3 O (2 O$_4$ and one O$_3$) ions close to the 
	Mn plane, and two O (O$_1$ and O$_2$) ions at the apexes. Corner sharing 
	MnO$_5$ bipyramids form layers separated along the c-axis by Yb layers.
        In the space group $P6_3cm$, Yb occupies the two crystallographically 
	inequivalent sites $2a$ and $4b$, with trigonal local symmetry around the 
	hexagonal c-axis. The Yb(4b) site consists of 4 equivalent atomic 
	positions inside the hexagonal unit cell, and the Yb(2a) site, of two 
	equivalent atomic positions along the c-axis-edges of the cell.
            \begin{table}
            \centering
                \begin{tabular}{|l|ccc|}
                   \hline
                   Atoms & x & y & z\\
                   \hline
                   \hline
                   Yb(2a) & 0 & 0 & 0.274(2)\\
                   Yb(4b) & $\frac{1}{3}$ & $\frac{2}{3}$ & 0.232(2)\\
                   Mn(6c) & 0.345(4) & 0 & 0\\
                   O$_1$(6c) & 0.307(2) & 0 & 0.165(3)\\
                   O$_2$(6c) & 0.640(1) & 0 & 0.336(3)\\
                   O$_3$(4b) & 0 & 0 & 0.475(2)\\
                   O$_4$(2a) & $\frac{1}{3}$ & $\frac{2}{3}$ & 0.020(2)\\
                   \hline
                   \hline
                   Discrepancy & Bragg R-factor & 4.32\% & \\
                   Factors & RF-factor & 3.21\% & \\
                   \hline
                \end{tabular}
                \caption{Atom positions and discrepancy factors at room 
	temperature}
                \label{position}
             \end{table}

    \section{Magnetic susceptibility and $^{170}$Yb M\"ossbauer data}
\begin{figure}
               \centering
               \includegraphics[width=9cm]{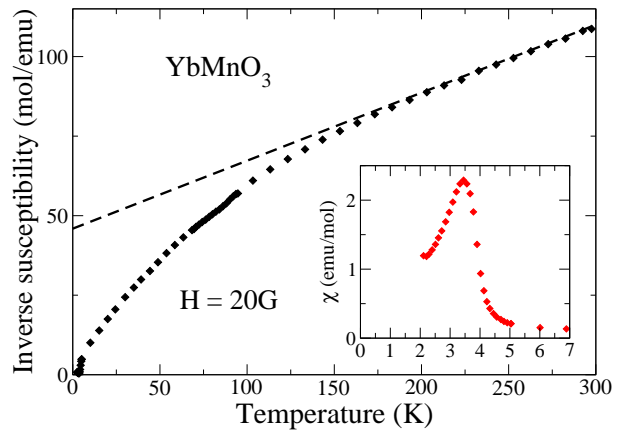}
               \caption{(Color online) Inverse  magnetic susceptibility in 
	YbMnO$_3$. The dashed line is a Curie-Weiss law with $\mu_{eff}$=6.1\,$\mu_B$
	and $\theta_p=-220$\,K. Insert: low temperature susceptibility.}
               \label{xi_20G}
           \end{figure}

        The thermal variation of the magnetic susceptibility $\chi(T)$, measured
        in a field of 20~G, is shown in Fig.\ref{xi_20G}. The inverse
        susceptibility follows a Curie-Weiss law in the temperature range
        200-300~K, with an effective moment $\mu_{eff}$ = 6.1(1)~$\mu_B$ and a
        paramagnetic Curie temperature $\theta_p \simeq - 220(5)$~K 
	(antiferromagnetic). As will be shown in section \ref{strumag}, the N\'eel
	temperature in YbMnO$_3$ is $T_N \simeq 85$\,K. The large negative $\theta_p$ value,
	such that $\vert \theta_p \vert/T_N \simeq 2.5$, could be linked with the 
	frustration of the Mn moments on their triangular lattice.
        Assuming for Yb$^{3+}$ the free ion value $\mu_{eff}$ = 4.54~$\mu_B$, we
	obtain an effective moment of 4.1(1)~$\mu_B$ for the Mn$^{3+}$ ion ($S$=2),
	which is lower than the value 4.9~$\mu_B$ expected for $g$=2. This deviation
	could be due to our limited experimental temperature range where the Curie-Weiss 
	law holds; indeed, susceptibility measurements performed in a single crystal 
	up to 350\,K \cite{fontcu_2008} obtain the correct effective moment with
	the free ion values for both ions.   
	 No anomaly in $\chi(T)$ is found at the magnetic transition, in contrast with 
	magnetic measurements in single crystals \cite{Katsu_2001,fontcu_2008}, 
	where $\chi_c(T)$ exhibits a small peak at $T_N$. At low temperature, 
	a ferromagnetic like increase of $\chi(T)$ occurs at 3.5~K (inset of 
	Fig.\ref{xi_20G}), signalling the onset of magnetic ordering among
        the Yb(2a) ions, as confirmed by the M\"ossbauer data to be described 
	next.
         \begin{figure}
             \centerline{ \resizebox{0.4\textwidth}{!}{
             \includegraphics{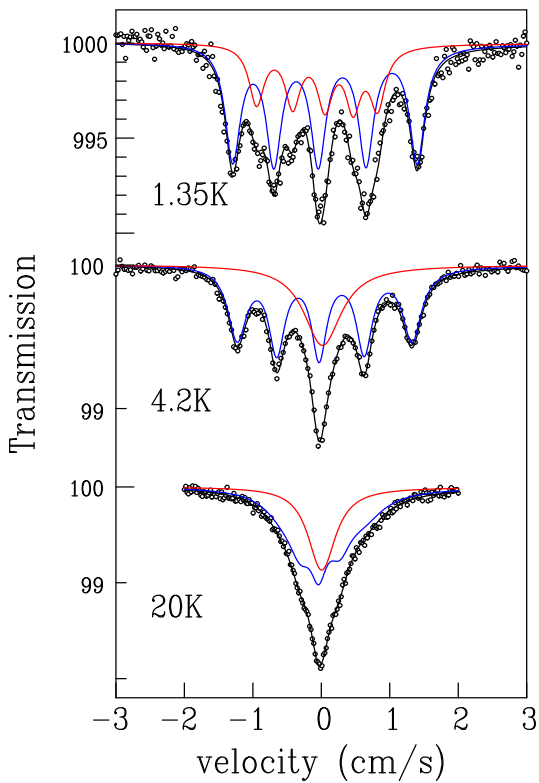}
             }}
             \caption{(Color online) $^{170}$Yb M\"ossbauer spectra in
             YbMnO$_3$ at selected temperatures. The subspectrum
             corresponding to the minority Yb(2a) site (relative occupancy 1/3), 
	     is drawn in red, that corresponding to the Yb(4b) site in blue.}
             \label{spe170}
         \end{figure}

        The $^{170}$Yb M\"ossbauer spectra at selected temperatures are shown in
        Fig.~\ref{spe170}. Between 1.35~K and 25\,K, they show two subspectra with
        relative weights close to the ratio 1:2, that we attribute
        respectively to Yb(2a) and Yb(4b) sites. Above 25\,K, the spectra are
        no longer resolved and consist of a single line. In Fig.~\ref{spe170}, the
        Yb(2a) subspectrum corresponds to the red line, the Yb(4b) to the blue
        line. At 1.35\,K, both subspectra are five-line magnetic hyperfine
        patterns, with a small quadrupolar hyperfine interaction. The
        weakness of the latter precludes any reliable information about
        the orientation of the hyperfine field with respect to the c-axis
        to be obtained. The hyperfine fields are respectively 117(3)\,T
        and 180(3)\,T for site Yb(2a) and Yb(4b). Using the
        proportionality constant $C$=102\,T/$\mu_B$ linking the hyperfine
	field and the Yb$^{3+}$ magnetic moment \cite{bonv0}, the saturated
        moment is 1.15(3)\,$\mu_B$ for Yb(2a) and 1.76(3)\,$\mu_B$ for
        Yb(4b).

        As temperature increases, the Yb(2a) hyperfine field, and thus
        the moment, slowly decreases down to 3~K, reaching 1.00(3)~$\mu_B$ at 
	3~K. A spectral change occurs between 3 and 3.5\,K on the Yb(2a) 
	subspectrum: the hyperfine field pattern transforms into a single 
	unresolved broad line at 3.5~K (see the red subspectrum at 4.2~K in
        Fig.~\ref{spe170}). Therefore, the hyperfine field vanishes on this site 
	above 3.5~K, and the Yb(2a) ion becomes paramagnetic. The M\"ossbauer 
	lineshape then reflects the ionic fluctuations within the ground Kramers 
	doublet, and its interpretation is postponed until the end of this section.
	The ordering of the Yb(2a) moments below 3.5~K is also observed on the 
	magnetic susceptibility, which shows a large ferromagnetic-like anomaly at 
	the transition (see insert of Fig.~\ref{xi_20G}).
         \begin{figure}
             \centerline{\resizebox{0.5\textwidth}{!}{
             \includegraphics{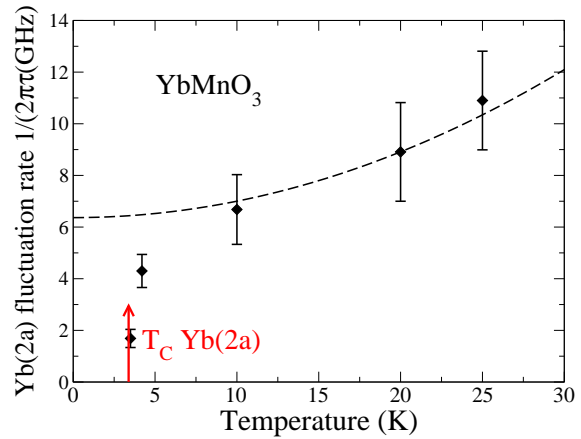}
             }}
             \caption{(Color online) Thermal variation of the Yb$^{3+}$(2a) 
	     relaxation rate extracted from the $^{170}$Yb M\"ossbauer spectra 
	     in YbMnO$_3$. The dashed line is a guide for the eye, and the red arrow
	     marks the (probably) ferromagnetic transition of the Yb(2a) sites.}
             \label{wrel}
         \end{figure}
        The Yb(4b) subspectrum shows no qualitative change from 1.35~K
        to 25~K: the hyperfine field decreases, reaching 59(5)~T (i.e. a
        moment of 0.58(5)~$\mu_B$) at 25~K. At higher temperature, it is
        not possible to distinguish the Yb(2a) and Yb(4b)
        sub-spectra. The thermal variation of the Yb(4b) moment is presented in
        section \ref{disc} in comparison with the moment values derived from 
	neutron diffraction. Our 4.2~K spectrum is in agreement with that of a 
	recent $^{57}$Fe and $^{170}$Yb M\"ossbauer study\cite{stewart_2008}.

        The behaviours of the Yb ions at the two sites are thus rather different:
	above 3.5~K, there is a vanishing molecular 
	field at the $2a$ site, which means that the exchange field from the Mn 
	ions is zero. This is not the case for the Yb ions at the $4a$ sites, 
	which are polarized by Mn exchange up to at least 20~K. The neutron study 
	described below shows that this exchange field is present up to 40~K, and 
	probably up to $T_N$. This behaviour is similar to that in HoMnO$_3$.

	The paramagnetic relaxation of the Yb(2a) ion can be interpreted
	using the M\"ossbauer relaxational lineshape developed in Ref.\onlinecite{gonz}. 
	Since the Yb(2a) subspectrum consists of a single line, the ``extreme narrowing''
	regime holds for the fluctuation rate $\frac{1}{2\pi\tau}$, i.e. $h/\tau \gg A$, where 
	$A$ is the hyperfine constant associated with the Yb$^{3+}$ ground Kramers  
	doublet and the excited nuclear state of the $^{170}$Yb isotope. It is related 
	to the above mentioned proportionality constant $C$ by \cite{abr}: $A = C\  g 
	\mu_B \ g_n \mu_n$, where $g$ and $g_n$ are respectively the electronic and
	nuclear gyromagnetic ratios (or g-factors), and $\mu_n$ the nuclear Bohr magneton.
	Using the Yb(2a) saturated moment value 1.15\,$\mu_B$, yielding $g$=2.3,
	one gets $A$ = 580\,MHz. Since the width of the Yb(2a) line narrows as 
	temperature increases, the spin relaxation rate increases, according to the 
	expression for the dynamical broadening \cite{gonz}: $\Delta \Gamma_R = 
	3 A^2 \tau$.  The fitted values of the relaxation rate $\frac{1}{2\pi\tau}$ are
	represented in Fig.\ref{wrel}. It slows down abruptly when approaching the transition
 	near 3.5\,K from above (``critical slowing down''); the $T=0$ ``extrapolated'' value of
	$\simeq$6\,GHz is due to exchange-driven spin-spin relaxation and the increase 
	above 10\,K can be caused by phonon driven relaxation.    

    \section{Magnetic Structure} \label{strumag}
\begin{figure}
                    \centering
                \includegraphics[width=8cm,height=5.5cm]{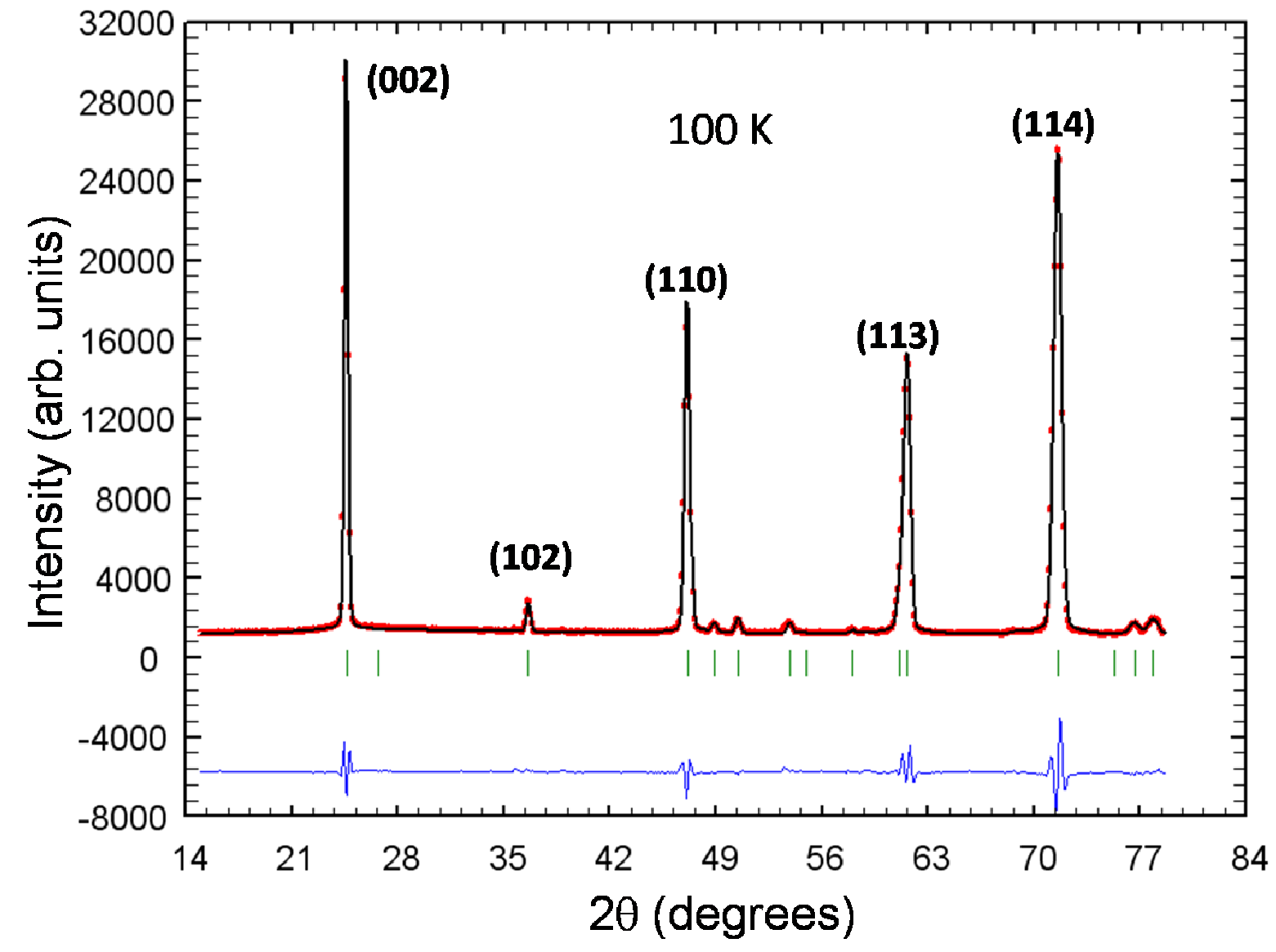}\\
                \includegraphics[width=8cm,height=5.5cm]{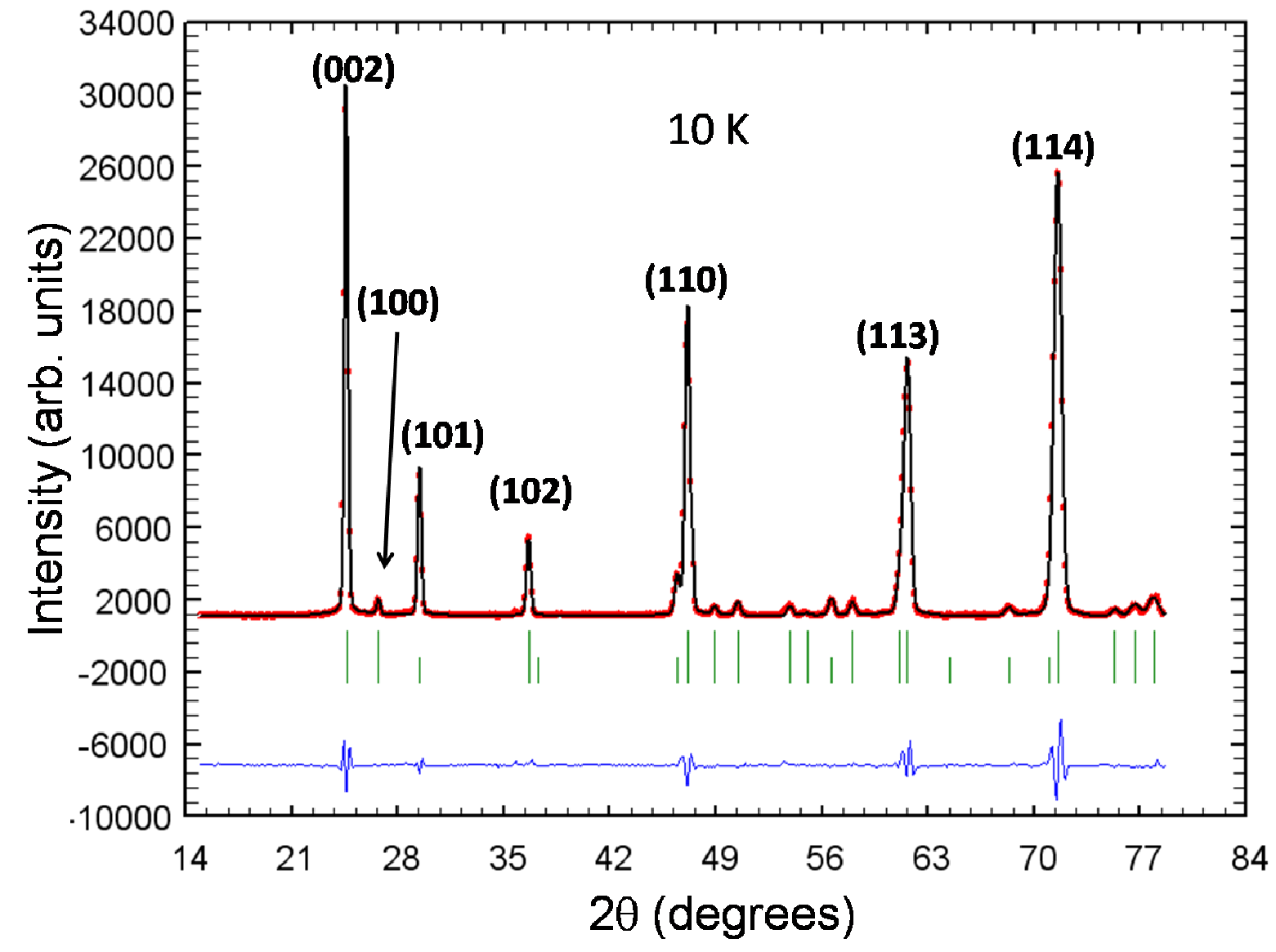}\\
                \includegraphics[width=8cm,height=5.5cm]{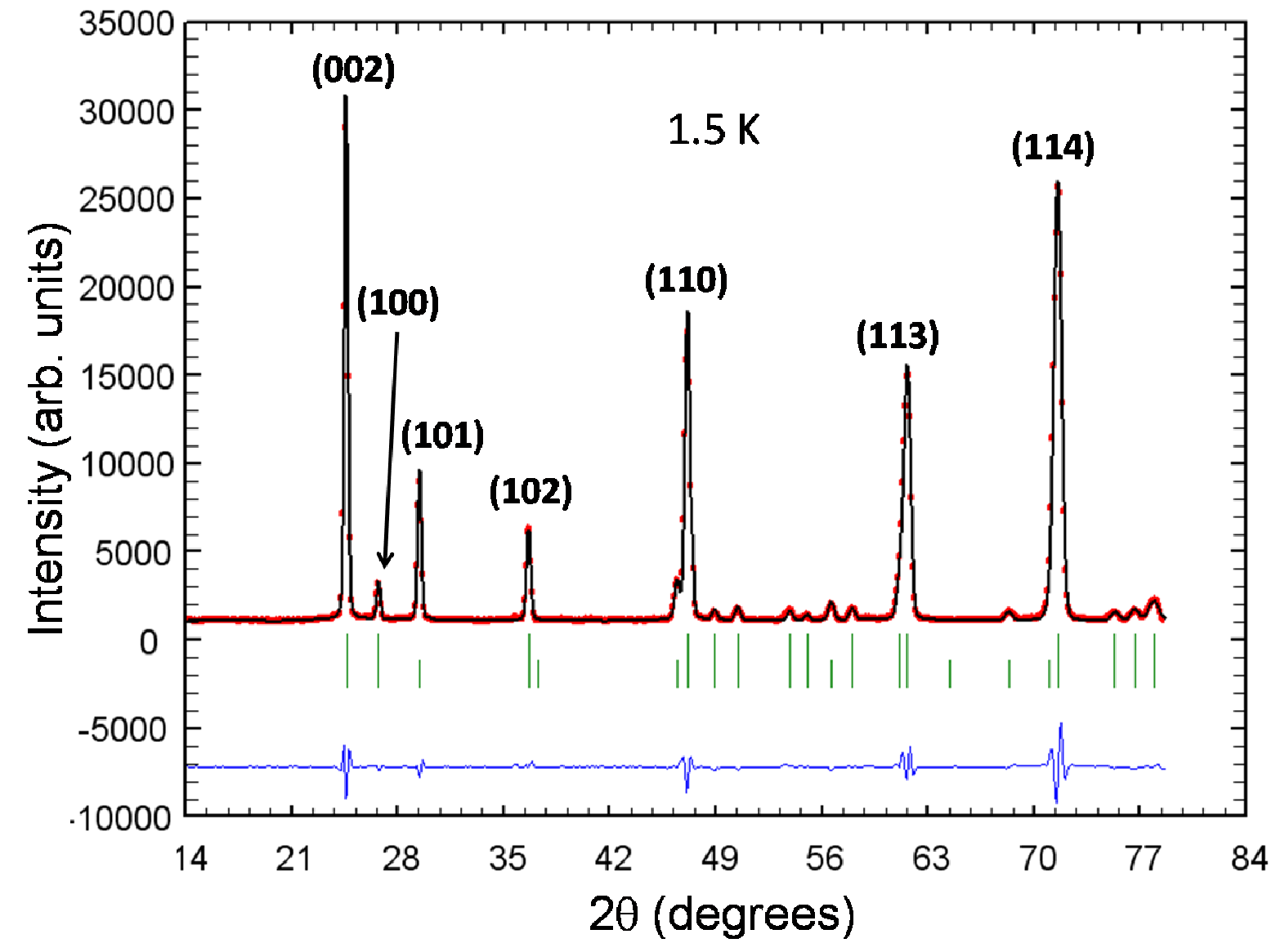}\\
                \caption{(Color online) Observed and Fullprof calculated NPD 
	patterns between 1.5~K and 100~K. The Bragg reflections (tics),
        and the difference between the observed and calculated patterns are 
	plotted at the bottom.}
                \label{NPD_temp}
            \end{figure}

\begin{figure}
                 \centering
                     \includegraphics[width=8cm,height=5.5cm]{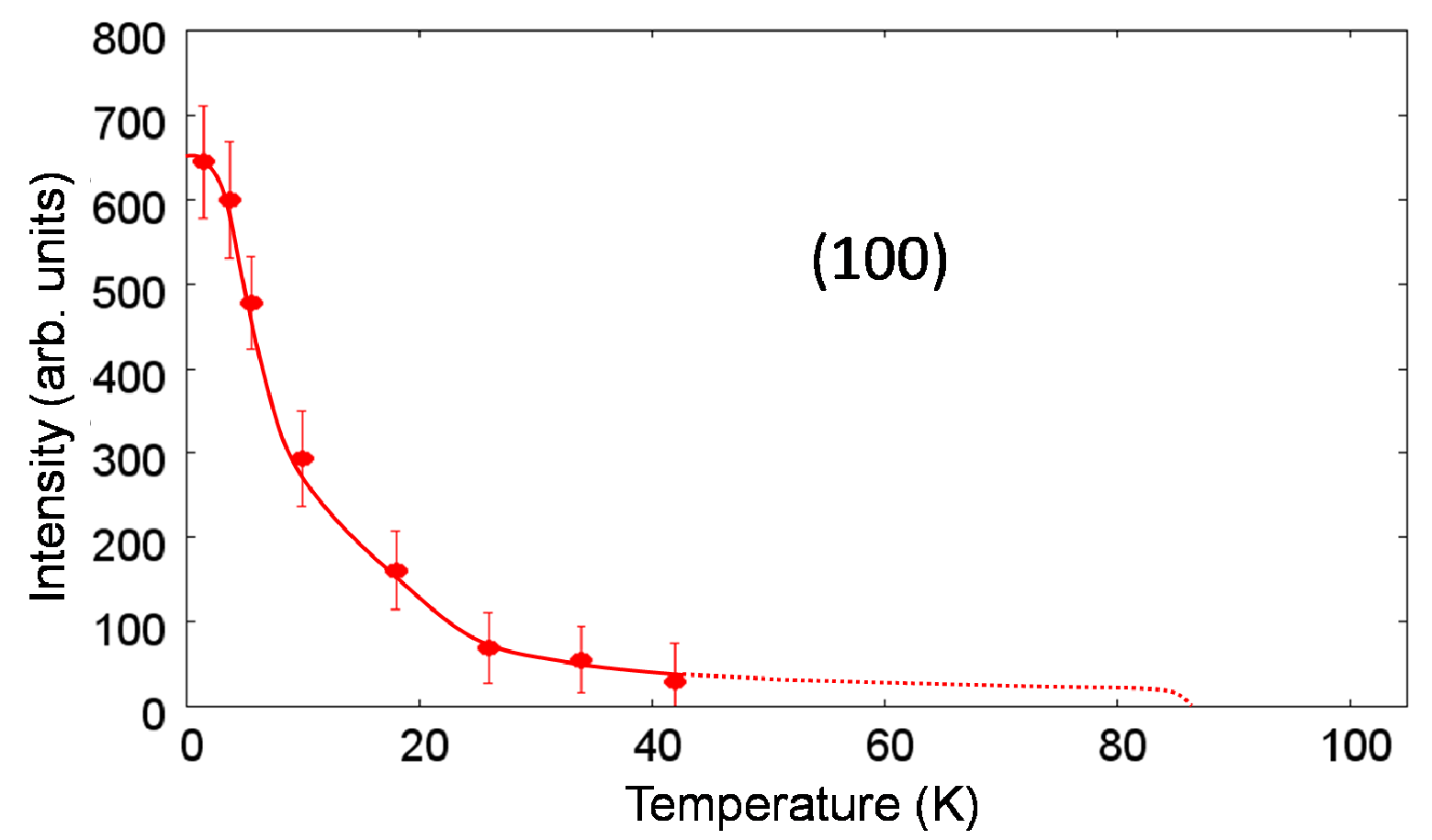}
                     \includegraphics[width=8cm,height=5.5cm]{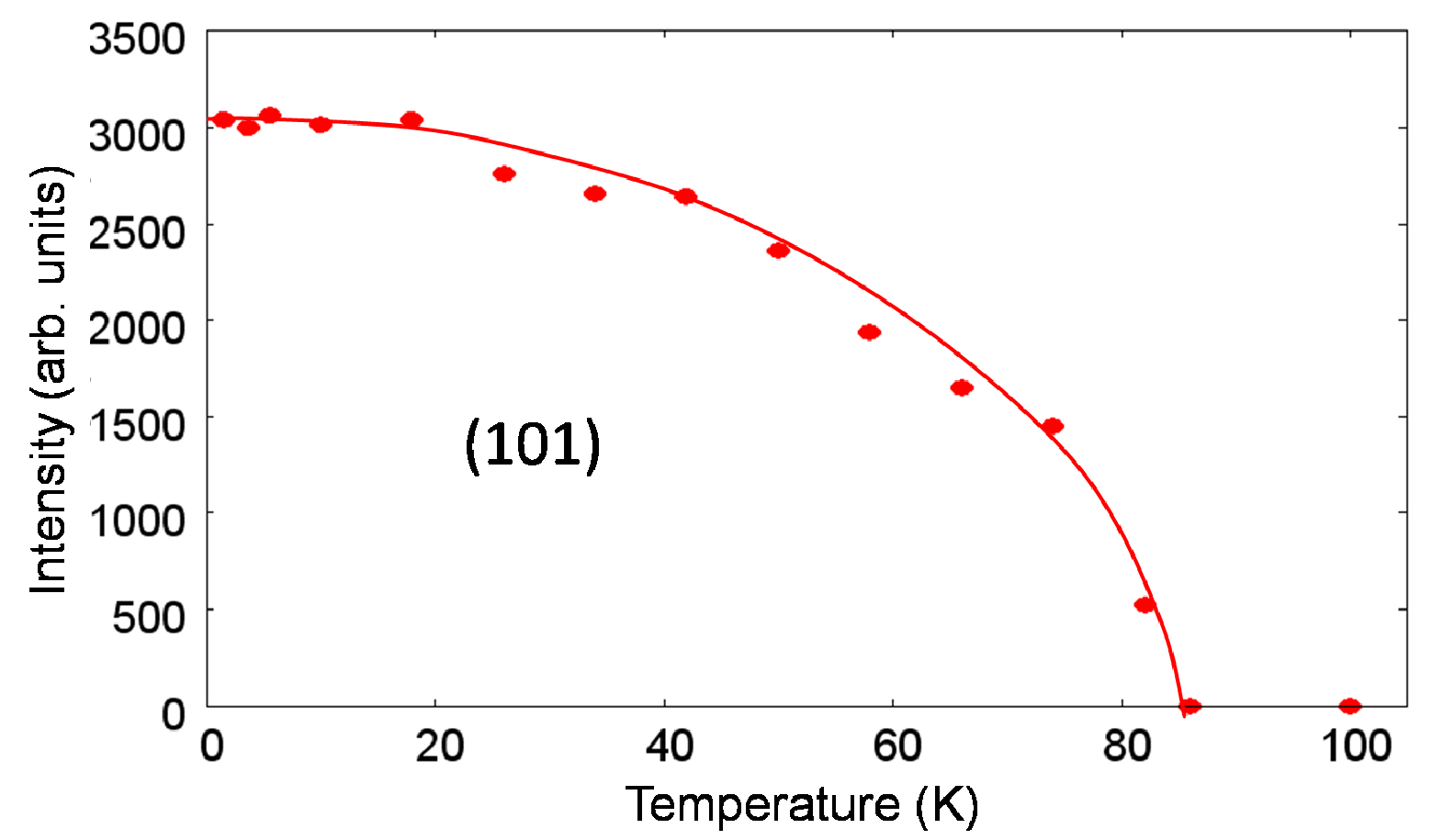}\\
                     \includegraphics[width=8cm,height=5.5cm]{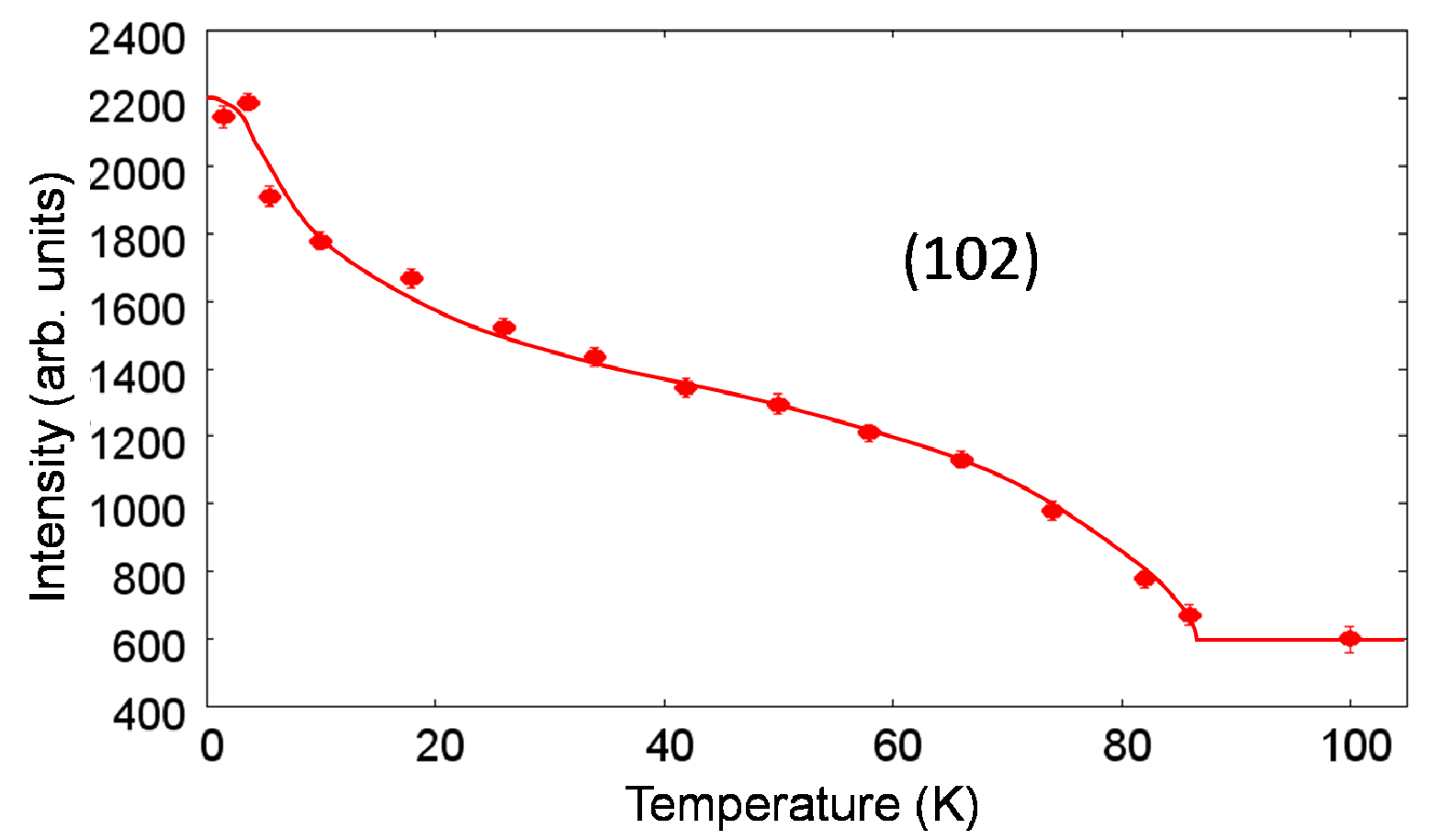}\\
                     \caption{(Color online) Integrated intensities of the (100), 
	(101) and (102) reflections between 1.5 and 100K (lines are guides to the 
	eyes).}
                     \label{Int_Int}
                 \end{figure}

        The thermal evolution of the NPD patterns is shown in 
	Fig.~\ref{NPD_temp}. The temperature dependences of the integrated 
	intensities of typical Bragg peaks (100), (101) and (102) are shown in 
	Fig.~\ref{Int_Int}. Strong  changes occur around T$_N=86~K$, which 
	corresponds to the magnetic transition. Below this temperature, an 
	intensity of magnetic origin grows at the (100) and (101) Bragg positions,
        where the nuclear reflections are forbidden by the $P 6_3 cm$ symmetry, 
	and the intensity of the (102) peak strongly increases due to an 
	additional magnetic contribution. All peaks can be indexed within the
        hexagonal space group $P 6_3 cm$ and a propagation vector $\vec{k}=
	\vec{0}$; this points to an AF structure having the same unit cell as 
	the chemical one, since a ferromagnetic ordering can be excluded from the
	absence of anomaly in the susceptibility at the transition.

        The temperature dependence of the (100) and (101) Bragg peaks strongly 
	differ from each other. The (100) peak remains rather weak down to about 
	20 K, then suddenly increases below. In contrast, the (101) peak increases
	abruptly below 86~K. These variations suggest that the (100) and (101) 
	peak intensities are controlled by Yb and Mn ordering, respectively, the 
	(102) Bragg peak involving contributions of both orderings. Moreover, one 
	observes a strong (101) peak in YbMnO$_3$, with thermal evolution akin to 
	that of the (100) peak in YMnO$_3$\cite{Munoz_2000}. Recalling that in 
	YMnO$_3$, the Mn order does not induce a strong (101) magnetic peak, it 
	means that Mn order is affected by Y/Yb substitution. Finally, the 
	temperature dependence of all magnetic peaks is monotonic, in contrast 
	with HoMnO$_3$ where Mn magnetic moments reorient at low temperature.

         \subsection{Mn order} \label{mnord}
\begin{figure}
                 \centering
                 \includegraphics[width=8cm]{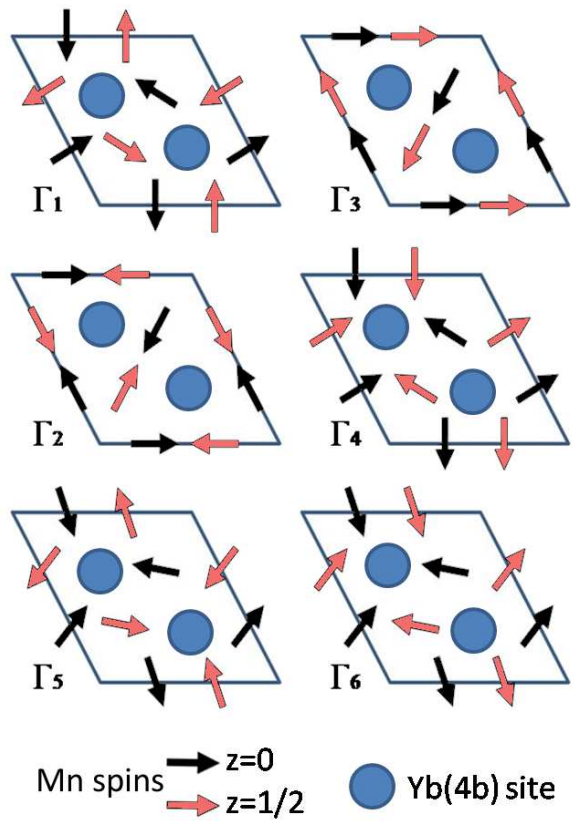}\\
                 \caption{(Color online) Symmetry allowed Mn spin orders in 
	hexagonal RMnO$_3$. In YbMnO$_3$, $\Gamma_2$ or $\Gamma_4$ IR yield the best 
	fit to experiment.}
                 \label{IR}
             \end{figure}

        To analyze the magnetic structure we searched for all irreductible
        representations (IR) compatible with the crystal symmetry using the theory
        of group representation analysis\cite{Bertaut_1961} and the program 
	Basireps\cite{Rod_Carv_2001}. Description of this analysis for hexagonal 
	RMnO$_3$ compounds was already given in Ref.\onlinecite{Munoz_2000}.
        The atomic positions in the unit cell were kept fixed and equal to those 
	determined above. In the space groupe $P6_3cm$ with $\vec{k}=\vec{0}$ 
	propagation vector, one finds 6 irreducible representations (IR), 
	corresponding to the 6 possible configurations of the Mn moments reported 
	in Fig.~\ref{IR}. In these configurations, the Mn moments lie in the (a,b)
	plane and their vectorial sum is zero. This results from the frustration 
	of the Mn moments in their triangular lattice.

        It is important to notice that the $\Gamma_2$ and the $\Gamma_4$ 
	configurations are homometric (namely cannot be distinguished by NPD)
        as well as $\Gamma_1$ and $\Gamma_3$\cite{Bertaut_IR_1963}.
        The best fit of the data was obtained assuming Mn ordering which 
	corresponds to the irreducible representation $\Gamma_2$ or $\Gamma_4$, 
	in contrast with YMnO$_3$ where it corresponds to a $\Gamma_1$ or 
	$\Gamma_3$ IR. The saturated Mn moment value is 3.25\,$\mu_B$, lower than
	the value 4\,$\mu_B$ expected for $g$=2 and $S$=2.

         \subsection{Yb order}
        The magnetic order of Yb moments is more intricate and difficult to 
	determine unambiguously using neutron diffraction only, since the two 
	crystallographic $2a$ and $4b$ sites order independently. Space group 
	symmetry allows two types of magnetic orders: along the c-axis or in the 
	(a,b) plane, and the coupling between the Yb ions can be antiferromagnetic
	or ferromagnetic (F), leading to six possibilities. The M\"ossbauer
        spectroscopy data show that the low temperature Yb magnetic moments are 
	small: 1.76~$\mu_B$ for Yb(4b) and 1.15~$\mu_B$ for Yb(2a), as compared 
	to the saturated Mn moment value of 3.25~$\mu_B$. Then the magnetic 
	ordering of the Yb  sublattices only gives incremental contributions to 
	the NPD spectra, and their magnetic structure was solved with the help of
	M\"ossbauer spectroscopy.

        In the temperature range 3.5~K$<$T$<$86~K, we considered contributions 
	only from ordered magnetic moments on the Yb(4b) sites. For T$<$3.5~K, we
        considered also ordered moments on the Yb(2a) sites. As to the $4b$
        sites, the best agreement with the NPD patterns is obtained for Yb moments
        oriented along the c-axis. We find that the Yb moments of the $4b$ sites 
	are antiferromagnetically coupled within a given layer, and  that the z 
	and z+$\frac{1}{2}$ layers are ferromagnetically  coupled. At $T=3.7~K$, 
	this refinement yields a discrepancy factor $R_{Mag}$ of about 3.1~\%, 
	much better than the values around 20~\% given by alternative solutions. 
	As shown in Fig.~\ref{Moment}, the moment values on the Yb(4b) sites 
	deduced from NPD patterns are in excellent agreement with those obtained 
	from the M\"ossbauer spectra.

        As  to the $2a$ sites, which order below 3.5~K with much smaller moments, 
	we fixed the moments to the values deduced from the M\"ossbauer spectra. The orientation
	of the moments was refined using our NPD datas. There are two possibilities for the Yb(2a)
	to orient : along the c-axis or in the (a,b) plane.
        Taking into account the ferromagnetic-like increase
	 of the susceptibility at 3.5~K, we assume a F coupling between two 
	adjacent Yb(2a) layers. We obtain a discrepancy factor of R=3.05~\% 
	for the $2a$ moments in the (a,b) plane, slightly better than for the 
	moments along the c-axis (R=3.44~\%). It suggests that the Yb(2a) moments 
	are perpendicular to the c-axis.

        Finally, the moment values at 1.5~K are $m_{Mn}$=$3.23(5)\,\mu_B$, 
	$m_{Yb(4b)}$=$1.77(5)\,\mu_B$ and $m_{Yb(2a)}$= $1.13(5)\,\mu_B$. The 
	evolution of the moments versus temperature is plotted in 
	Fig.~\ref{Moment}.

    \section{Discussion} \label{disc}
        In YbMnO$_3$, the Mn magnetic order with $\Gamma_2$ or $\Gamma_4$ symmetry
        agrees with expectations from optical measurements\cite{Raman}. Such a
        type of order is also expected from the general tendency observed in
        hexagonal RMnO$_3$ compounds versus the rare earth ionic radius
        \cite{Kozlenko_2007}.
        $\Gamma_{2-4}$ magnetic orders seem to be stabilized for low ionic radius
        (Er,Yb,Lu), and $\Gamma_{1-3}$ for higher ionic radius (Ho, Y, Y-Er).
        This could be related to the amplitude of the distortion of the Mn lattice
        with respect to triangular symmetry.

        As  concerns the rare earth moments, we found that the $4b$ and $2a$ sites
        order independently. This is also the case in HoMnO$_3$, where ordered Ho
        moments on $4b$ sites are observed below 35-40~K, whereas Ho $2a$
        sites remain paramagnetic down to 5~K\cite{Lynn}. In HoMnO$_3$,
        a change of the Mn structure (from $\Gamma_{2-4}$  above 38~K to
        $\Gamma_{1-3}$ below) occurs together with the rare earth ordering. This 
	is not the case for YbMnO$_3$, possibly due to the lower value of the Yb 
	moment.

        In spite of a very large amount of experimental and theoretical work, the 
	nature of the interactions which control the magnetic structure in 
	RMnO$_3$ compounds is still unclear\cite{Fiebig_2002,Lynn}. To shed some 
	light on this point, we first compare the measured temperature dependence 
	of Mn and Yb moments with calculations from a molecular field model. Then 
	we briefly discuss the symmetry of interactions necessary to explain the
        observed magnetic orders. Finally we propose possible origins for the
        magnetic interactions in RMnO$_3$, compatible with both experiment and
        calculations.

        \subsection{Molecular field analysis}

\begin{figure}
                    \centering
                    \includegraphics[width=8cm]{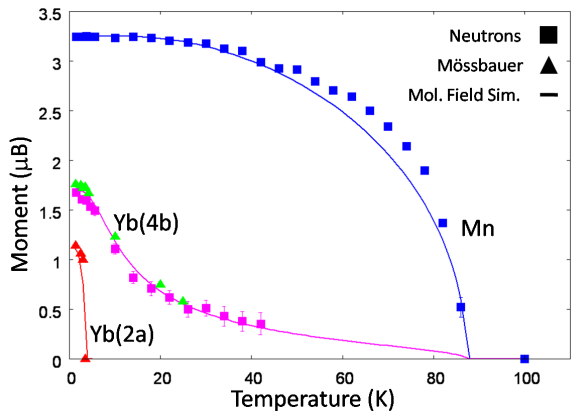}\\
                    \caption{(Color online) Observed Mn, Yb(4b) and Yb(2a) moments
        from NPD and M\"ossbauer measurements (dots) versus temperature. The solid
        lines correspond to the molecular field calculation.}
                    \label{Moment}
                \end{figure}

        The thermal variations of the Mn and Yb magnetic moments derived from
        both neutron and M\"ossbauer data are shown in Fig.~\ref{Moment}. The
        Yb(4b) moment values derived from both techniques are in remarkable
        agreement, and their thermal variation has an unusual shape, with upwards
        curvature in a large temperature range. We applied the molecular field 
	model in order to calculate these thermal variations. We first obtained 
	the $T$ dependence of the Mn moment by a self consistent mean field 
	calculation using the Brillouin function $B_2(x)$ for S=2 and a
        molecular field constant $\lambda_0$ representing Mn-Mn exchange. The
        standard formula has to be slightly modified because the saturated Mn 
	moment of $m_{sat}$=3.23(5)~$\mu_B$ is lower than $gS\mu_B$=4~$\mu_B$. So we used 
	the following expression:
             \begin{equation}
                m_{Mn}(T) = m_{sat}\ B_2\left[\frac{g \mu_B S \ 
	\lambda_0 m_{Mn}(T)}{k_B T}\right],
                \label{self}
             \end{equation}
        which ensures that saturation occurs at $m_{sat}$=3.3~$\mu_B$. We obtain 
	a good fit of the experimental data  (blue line in Fig.~\ref{Moment}) 
	with $\lambda_0$=19~T/$\mu_B$. Then we fitted the $T$ dependence of the 
	Yb(4b) moment assuming that the ground crystal field doublet alone is 
	populated, i.e. describing the Yb ion by an effective spin $S$=1/2, with 
	a g-factor $g_{Yb}$ in the direction of the net exchange field from the 
	Mn ions. Since the saturated Yb(4b) moment is $m_0 = \frac{1}{2} g_{Yb} 
	\mu_B$ = 1.75~$\mu_B$, then: $g_{Yb}$=3.5. The Yb(4b) ion is submitted to 
	a molecular field $\lambda_1 m_{Mn}$ from Yb-Mn exchange, since we 
	neglect Yb-Yb exchange. Its moment is then calculated using the 
	expression:
             \begin{equation}
                m_{Yb}(T) = \frac{1}{2} g_{Yb} \mu_B \ \tanh\left[\frac{g_{Yb} 
	\mu_B \ \lambda_1 m_{Mn}(T)}{2k_B T}\right],
             \end{equation}
        where $m_{Mn}(T)$ is obtained from expression (\ref{self}). We obtain a 
	good fit to the data (magenta line in Fig.~\ref{Moment}) with $\lambda_1$ =
        2.1~T/$\mu_B$.  The unusual upwards curvature comes from the $\tanh$
        function with a small argument, the constant $\lambda_1$ being much 
	smaller than $\lambda_0$. With this model, a small ordered Yb moment 
	should persist up to T$_N$=~86K, with a value below 0.3~$\mu_B$, beyond 
	the accuracy of neutron and M\"ossbauer probes. We conclude that the Yb 
	ions on the $4b$ sites orient in a non-vanishing net molecular field 
	through Yb-Mn exchange. As  to the Yb moment on the $2a$ sites, their 
	much lower ordering temperature (3.5~K) strongly suggests that they 
	orient in their own molecular field, through Yb-Yb interactions. From 
	this analysis follows the hierarchy of exchange interactions in YbMnO$_3$,
	 in decreasing order : Mn-Mn, Yb-Mn and Yb-Yb. Similar conclusions could 
	be valid for other R ions (Ho, Er) in hexagonal RMnO$_3$.

        \subsection{Mn environment of a R ion}
        Assuming that R(4b) moments order via R-Mn interactions, we consider the 
	Mn environment around a R(4b) site, taking into account nearest neighbor 
	Mn ions only. In the following we will consider only the first four IR 
	($\Gamma_1$ to $\Gamma_4$) which describe the magnetism of the whole 
	RMnO$_3$ family (see fig 5). The moment on a R(4b) site is expected to 
	orient along the c-axis, to obey the trigonal symmetry of its Mn 
	environment. This agrees with our observations in YbMnO$_3$ and with the 
	magnetic order in HoMnO$_3$\cite{Munoz_2001, Brown_2006}.

        In order to depict the local magnetic environment of each R(4b) site, we 
	need to take into account the orientation and position of the Mn moments.
        For this purpose, we define the A factors as :
             \begin{eqnarray}
                A_{i} & = & \sum_{j}\vec{S}_{j}.\vec{r}_{ij},
             \end{eqnarray}
        where the sum runs over the three Mn first neighbors of the R(4b) site in 
	a particular plane, $\vec{S}_{j}$ is the magnetic moment of Mn at site $j$
	and $\vec{r}_{ij}$ is the vector joining Mn site $j$ to R(4b) site $i$.

        The A factors are either positive (+) or negative ($-$) for a given Mn 
	plane depending of the magnetic configuration $\Gamma_i$ of the Mn ions.
        This leads to four possible magnetic environments for the R(4b) ion, 
	(+ +), ($-$ $-$), (+ $-$) and ($-$ +), defined by the signs of the 
	(A$_{i}^{up}$ A$_{i}^{down}$) pairs associated with upper and lower Mn
        planes. One can easily see that there is either only one or two opposite 
	A-pairs allowed for each IR, whatever the IR retained for the Mn order.
        Considering that R moments in environments with identical (resp. opposite)
	A-pairs align parallel (resp. antiparallel) along c, one can predict the 
	R magnetic order on a 4b site for a given Mn order. This is reported in
        Table \ref{sym}.
             \begin{table}
                \begin{tabular}{| m{1.5cm} |  c  | c c |}
                    \hline
                    IR & \multirow{2}{*}{A-pair} & \multicolumn{2}{c|}{R(4b) coupling} \\
                       & & Intraplane & Interplane \\
                    \hline
                    $\Gamma_1$  & (+ $-$) or ($-$ +) & AF & AF\\
                    $\Gamma_2$  & ($-$ $-$) & F & F\\
                    $\Gamma_3$  & ($-$ +) or (+ $-$) & F & AF\\
                    $\Gamma_4$  & (+ +) or ($-$ $-$) & AF & F\\
                    \hline
                \end{tabular}
                \caption{R(4b) magnetic orders deduced for each configuration 
	$\Gamma_i$ of the Mn moments: i) by our analysis of the local magnetic 
	environment using the sign of the A-pairs; ii) by the minimization of DM 
	and dipolar energies. Both methods yield the same result.}
                \label{sym}
             \end{table}
        This description provides a simple way to distinguish two homometric 
	configurations, knowing the order on the 4b sites. Experimentally,
        the intraplane Yb(4b) coupling is AF while the interplane coupling is F. 
	Therefore, the Mn order must be of $\Gamma_4$ type and $\Gamma_2$ can be 
	discarded. This description also agrees with recent experiments
	\cite{Munoz_2001, Nandi} in HoMnO$_3$ and can be easily extended to the 
	rest of the RMnO$_3$ family.

        \subsection{Origin of the molecular field}
        The exchange energy between a Yb ion and its Mn first neighbors is 
	classically written as a sum over Mn sites $\sum J\vec{S}_{Mn}.
	\vec{S}_{Yb}$, where $J$ is the isotropic first neighbor exchange 
	interaction. Each Yb atom (either on a $2a$ or $4b$ site) is at the top 
	of a pyramid with an equilateral triangular basis of Mn ions.
        Therefore, 
        the exchange field induced at either Yb site cancels by symmetry for any 
	of the $\Gamma_{1-4}$ representations of the Mn magnetic order. Moreover, 
	the classical exchange term cannot explain the perpendicular orientations 
	of the R and Mn moments.

        Experimentally, the exchange field is effectively zero at the Yb(2a) site 
	(above 3.5~K), but it is not at the Yb(4b) site. So, in order to explain 
	the R(4b) magnetic order, we have to introduce an interaction which is 
	not isotropic, i.e. which depends on $\vec{r}_{ij}$, the radius vector
        linking a R(4b) to a Mn ion. This is the only way to obtain an out of 
	plane effective field.

        In the following we consider two interactions fulfilling this condition: 
	the magnetic dipolar interaction and the Dzyaloshinskii-Moriya (DM) 
	coupling between R(4b) and Mn spins. The magnetic dipolar interaction is 
	defined as :
            \begin{eqnarray}
                {\cal H}_{dip} & = & - \sum_{i,j} \vec{\sigma}_i\vec{B}_{ij},
            \end{eqnarray}
        where $\vec{B}_{ij}$ is the dipolar field induced by Mn moment $j$ at 
	R(4b) site $i$, and $\vec{\sigma}_i$ the spin of the R(4b) ion at site 
	$i$.

        The Dzyaloshinskii-Moriya (DM) interaction between the Yb(4b) and the Mn 
	moments is written as
             \begin{eqnarray}
             {\cal H}_{DM} & = & \sum_{i,j} \vec{D}_{ij}.(\vec{\sigma}_i \times\vec{S}_{j}).
            \end{eqnarray}
         According to the crystal symmetry and the rules defined by Moriya 
	\cite{Moriya}, $\vec{D}_{ij}$ is perpendicular to the mirror plane 
	including both R and Mn sites, i.e. to the c-axis, and defined as 
	$\vec{D}_{ij} = \eta \vec{c} \times \vec{r}_{ij}$, where $\eta$ is a 
	spin-orbit coupling constant.

        Although the dipolar and DM interactions are intrinsically different, 
	they may be expressed in similar ways in terms of the A factors:
            \begin{eqnarray}
                {\cal H}_{dip} & = & \sum_{i}\sigma_i  [(A_i\ r^z)^{up}+(A_i\ r^z)^{down}]\\
                {\cal H}_{DM}  & = & \eta \sum_{i}\sigma_i(A_i^{up}+A_i^{down}).
            \end{eqnarray}
        Minimizing the energy for these interactions should therefore lead to 
	the same configurations as found in section B. To confirm this, we have 
	rewritten both interactions in terms of a matrix $J_{ij}$ 
	\cite{Fiebig_2002} such as ${\cal H}=\sum_{i,j}\vec{\sigma}_i J_{ij} 
	\vec{S}_j$, where $J_{ij}$ is expressed as:
            \begin{eqnarray}
                J_{ij}^{DM} & = & \begin{pmatrix}
                0 & D_{ij}^z & -D_{ij}^y\\
                -D_{ij}^z & 0 & D_{ij}^x\\
                D_{ij}^y & -D_{ij}^x & 0
                \end{pmatrix} \\
                J_{ij}^{dip.} & = & \begin{pmatrix}
                r_{ij}^x r_{ij}^x & r_{ij}^x r_{ij}^y & r_{ij}^x r_{ij}^z\\
                r_{ij}^y r_{ij}^x & r_{ij}^y r_{ij}^y & r_{ij}^y r_{ij}^z\\
                r_{ij}^z r_{ij}^x & r_{ij}^z r_{ij}^y & r_{ij}^z r_{ij}^z
                \end{pmatrix}\frac{1}{r_{ij}^5}.
            \end{eqnarray}
        It is worth noting that the above matrixes both have non-diagonal
        terms, which depend on the R-Mn pair considered. This is a necessary 
	condition to ensure a non-zero interaction between Mn moments in (a,b) 
	planes and R(4b) moments along the c-axis.

        We determined the stable Yb(4b) configuration for each IR of the Mn order.
        We find that both dipolar and DM interactions give the same result, in 
	agreement with our analysis of the magnetic environment made above (see 
	Table \ref{sym}). Namely they stabilize the $\Gamma_4$ configuration of 
	the Mn moments for the Yb(4b) order determined experimentally.

        The analysis described here provides the conditions required for the 
	magnetic interactions to yield the R and Mn magnetic orders observed 
	experimentally. To go further, a quantitative calculation of the energy 
	should be made. Our estimations made for the the dipolar interaction show 
	that it is too weak by one order of magnitude to fully account for the 
	molecular field constant $\lambda_{Mn-Yb}$(=$\lambda_{1}$) = $
	2.1~T\mu_B^{-1}$ found in section A, Fig.~\ref{Moment}. As for the DM 
	interaction, an estimation would require an evaluation of the relevant 
	spin-orbit coupling constant.

        We hope that this analysis will stimulate further theoretical work
        on the R-Mn interactions at the origin of the molecular field. This will 
	allow one to understand the microscopic origin of the rare earth magnetic 
	order in the hexagonal RMnO$_3$ family.\\

    \section{Conclusion}
        We have studied the magnetic order in YbMnO$_3$ and found a consistent
        description by combining M\"ossbauer and neutrons probes. The temperature 
	dependences of the Yb and Mn magnetic moments are well fitted in a 
	molecular field approach, showing that  Yb(4b) moments order due to the 
	Mn molecular field, whereas the Yb(2a) moments order at much lower
        temperature through Yb-Yb interactions.

        We could determine the Mn magnetic structure and distinguish between two 
	homometric configurations by considerations about the magnetic 
	environment of each R(4b) sites. In YbMnO$_3$, we found a $\Gamma_4$
        configuration for Mn moments associated with F stacking of AF Yb(4b) 
	layers. This approach can be extended to the whole hexagonal RMnO$_3$ 
	family. We propose possible mechanisms for the R-Mn
        interactions to be studied by further theories.

        \textbf{Aknowledgements :} We thank D. Colson for the sample preparation, 
	F. Porcher and F. Bour\'ee for the measurement on 3T2.


\begin{thebibliography}{}
        \bibitem {Fiebig_revue} M. Fiebig, J. Phys. D : Appl. Phys., \textbf{38}, 
	R123-R152 (2005)

        \bibitem {Cheong_Nature} S.W. Cheong, M. Mostovoy, Nature, \textbf{6}, 
	13-20 (2007)

        \bibitem {Park_Y} J. Park, J.G. Park, Gun Sang Jeon, Han-Yong Choi, Changhee 
	Lee, W. Jo, R. Bewley, K. A. McEwen, T. G. Perring , Phys. Rev. B, \textbf{68}, 
	104426 (2003)

        \bibitem {Lee} S. Lee, A. Pirogov, M. Kang, K.H. Jang, M. Yonemura, T. Kamiyama, 
	S.-W. Cheong, F. Gozzo, N. Shin, H. Kimura, Y. Noda, J.G. Park, Nature, 
	\textbf{451}, 06507 (2008)
        
	\bibitem {Pimenov}A. Pimenov, T. Rudolf, F. Mayr, A. Loidl, A. A. Mukhin, 
	A. M. Balbashov, Phys. Rev. B, \textbf{74}, 100403(R) (2006)

        \bibitem {Katsura}H. Katsura, A. V. Balatsky, N. Nagaosa, Phys. Rev. Lett., 
	\textbf{98}, 027203 (2007)

        \bibitem {Sylvain}S. Petit, F. Moussa, M. Hennion, S. Pailh\`es, 
	L. Pinsard-Gaudart, A. Ivanov, Phys. Rev. Lett., \textbf{99}, 266604 (2007)

        \bibitem {Kimura}T. Kimura, G. Lawes, T. Goto, Y. Tokura, A. P. Ramirez, 
	Phys. Rev. B, \textbf{71}, 224425 (2005)

        \bibitem {Huang_1997} Z. J. Huang, Y. Cao, Y. Y. Sun, Y. Y. Xue, C. W. Chu, 
	Phys. Rev. B, \textbf{56}, 2623 (1997)

        \bibitem {Munoz} E. F. Bertaut, Physics Letters, \textbf{5}, 27-29 (1963)

        \bibitem {Raman} M. Fiebig, D. Frohlich, K. Kohn, S. Leute, T. Lottermoser, 
	V. V. Pavlov, R. V. Pisarev, Phys. Rev. Lett., \textbf{84}, 5620 (2000)

        \bibitem {Alonso_2000} J. A. Alonso, M.J. Martinez-Lope, M.T. Casais, 
	M.T. Fernandez-Diaz, Inorg. Chemistry, \textbf{39}, 917-923 (2000)

        \bibitem {Rod_Carv_1993} J. Rodriguez-Carvajal, Physica B, \textbf{192}, 
	55-69 (1993)

        \bibitem {Isobe_1991} M. Isobe, N. Kimizuka, M. Nakamura, T. Mohri, Acta 
	Cryst. Sect. C, \textbf{47}, 423-424 (1991)

        \bibitem {Van_2001}  B. B. Van Aken, A. Meetsma, T. T. M. Palstra, Acta 
	Cryst. Sect. E, \textbf{57}, i87-i89 (2001).

        \bibitem {Munoz_2000} A. Munoz, J. A. Alonso, J. Martinez-Lope, T. Casais, 
	J. L. Martinez, M. T. Fernandez-Diaz, Phys. Rev. B, \textbf{62}, 9498 (2000)

        \bibitem {fontcu_2008} J. Fontcuberta, M. Gospodinov, V. Skumryev, J. App. 
	Phys., \textbf{103}, 07B722 (2008)

        \bibitem {Katsu_2001} T. Katsufuji, S. Mori, M. Masaki, Y. Moritomo, 
	N. Yamamoto, H. Takagi, Phys. Rev. B, \textbf{64},104419 (2001)

        \bibitem {bonv0} P. Bonville, P. Imbert, G. J\'ehanno, F. Gonzalez-Jimenez, 
	F. Hartmann-Boutron, Phys. Rev. B, \textbf{30}, 3672 (1984).

        \bibitem {gonz} F. Gonzalez-Jimenez, P. Imbert, F. Hartmann-Boutron, 
	Phys. Rev. B \textbf{9} 95 (1974).

	\bibitem{abr} A. Abragam and B. Bleaney, {\it Electron Paramagnetic Resonance
	of Transition Ions} (Clarendon, Oxford, 1969)

        \bibitem {stewart_2008} H. A. Salama, G. A. Stewart, D. H. Ryan, 
	M. Elouneg-Jamroz, A. V. J. Edge, J. Phys. : Condens. Matter, \textbf{20}, 
	255213 (2008)

        \bibitem {Bertaut_1961} E. F. Bertaut, C. R. Hebd. S\'eances Acad. Sci., 
	\textbf{252},76 (1961); J. Phys. Radium, \textbf{22}, 321 (1961).

        \bibitem {Rod_Carv_2001} J. Rodriguez-Carvajal,
        http://www.ill.eu/sites/fullprof/php/programsfa7c.html?pagina=GBasireps 

        \bibitem {Bertaut_IR_1963} E. F. Bertaut, edited by G.T. Rado and H. Shul 
	(Academic, New York), Vol. 3, Chap. 4, p. 149. (1963)

        \bibitem {Kozlenko_2007} D. P. Kozlenko, S. E. Kichanov, S. Lee, J. G. Park, 
	B. N. Savenko, J. Phys. : Condens. Matter, \textbf{19}, 156228 (2007)

        \bibitem {Fiebig_2002} M. Fiebig, C. Degenhardt, R. V. Pisarev, 
	Phys. Rev. Lett., \textbf{88}, 027203 (2001)

        \bibitem {Lynn} O. P. Vajk, M. Kenzelmann, J.W. Lynn, S.B. Kim, 
	S.-W. Cheong, Phys. Rev. Lett., \textbf{94}, 087601 (2005)

        \bibitem {Munoz_2001} A. Munoz, J.A. Alonso, M.J. Martinez-Lope, M.T. Casais, 
	J.L. Martinez, M.T. Fernandez-Diaz,  Chem. Matters, \textbf{13}, 1497-1505 (2001)

        \bibitem {Brown_2006} P.J. Brown, T. Chatterji, J. Phys. : Condens. Matter, 
	\textbf{18}, 10085-10096 (2006)

        \bibitem {Nandi} S. Nandi, A. Kreyssig, L. Tan, J.W. Kim, J.Q. Yan, J.C. Lang, 
	D. Haskel, R. J. McQueeney, A. I. Goldman, Phys. Rev. Lett., \textbf{100}, 
	217201 (2008)

        \bibitem {Moriya} T. Moriya, Phys. Rev., \textbf{120}, 91 (1960)

    \end{thebibliography}
\end{document}